\documentclass[aps,prb,reprint,superscriptaddress,showpacs]{revtex4-1}

\usepackage[utf8]{inputenc}
\usepackage[english]{babel}
\usepackage{mathrsfs}
\usepackage{bm}
\usepackage{amssymb}
\usepackage{graphicx}
\usepackage{amsmath}
\usepackage{subeqnarray}
\usepackage{color}

\begin{document}

\title{Coupling magneto-elastic Lagrangians to spin transfer torque sources}

\author{Thomas Nussle}
\email{thomas.nussle@cea.fr}
\affiliation{CEA DAM/Le Ripault, BP 16, F-37260, Monts, FRANCE}
\affiliation{CNRS-Institut Denis Poisson (UMR7013), Université de Tours, Université d'Orléans, Parc de Grandmont, F-37200, Tours, FRANCE}

\author{Pascal Thibaudeau}
\email{pascal.thibaudeau@cea.fr}
\affiliation{CEA DAM/Le Ripault, BP 16, F-37260, Monts, FRANCE}

\author{Stam Nicolis}
\email{stam.nicolis@lmpt.univ-tours.fr}
\affiliation{CNRS-Laboratoire de Mathématiques et Physique Théorique (UMR 7350), Fédération de Recherche "Denis Poisson" (FR2964), Département de Physique, Université de Tours, Parc de Grandmont, F-37200, Tours, FRANCE}

\date{\today}

\begin{abstract}
	The consequences of coupling magnetic and elastic degrees of freedom, where spins and deformations are carried by point-like objects subject to local interactions, are studied, theoretically and by detailed numerical simulations. From the constrained Lagrangians we derive consistent equations of motion for the coupled dynamical variables. In order to probe the dynamics of such a system, we consider external perturbations, such as spin transfer torques for the magnetic part, and homogeneous stresses for the elastic part, associated to their corresponding damping. This approach is applied to the study of ultrafast switching processes  in  anti-ferromagnetic systems, which have recently attracted attention as  candidates for anti-ferromagnetic spintronic devices. Our strategy is then checked in  simple, but instructive, situations. We carried out numerical experiments to study, in particular,  how the magnetostrictive coupling and external stresses affect the nature of the switching processes  in  a prototype  anti-ferromagnetic material.
\end{abstract}


\maketitle

\section{Introduction}
The simplest classical field theory to describe the consequences of local interactions between magnetic and mechanical degrees of freedom is set up and its consequences are studied by numerical methods.

The starting point is a single, point-like, object carrying both, a classical spin vector, and a mechanical strain tensor, which can both depend on  time. Early attempts  may be found in many references \cite{rado_magnetism:_1963,cracknell_magnetoelastic_1974,suhl_relaxation_2007}.

In the canonical formulation, one has to consider the Lagrangian functional density $\mathscr{L}$ as a sum of three  main contributions:
The first one is the magnetic part, labeled ${\mathscr{L}}_\mathrm{s}$,  a functional of both a vector ${\bm{s}(t)}$ and its velocity $\dot{\bm{s}}(t)$.
Here the classical spin (or magnetic moment), i.e. the vector $\bm{\mu}(t)$, is to be identified with $\dot{\bm{s}}(t)$ instead of $\bm{s}(t)$ \cite{brink_lagrangian_1977}.

This can be explained as follows~:
As there is no point--like ``magnetic charge", in order to deduce an equation for the spin precession, that is second order in time,  the potential vector has to depend on the history of the variable $\bm{s}(t)$, hence it is non-locally dependent on it.
Another point of view would be to consider a ``magnetic monopole", but such considerations, that lead to so many implications beyond the classical level of description we want to address, will not be discussed here \cite{shnir_magnetic_2005}.

The second one is the mechanical part, labeled ${\mathscr{L}}_\mathrm{m}$,  a functional of the  symmetric Cauchy strain tensor ${\epsilon}_{ij}(t)$ and its time derivative ${\dot\epsilon}_{ij}(t)$. It represents a first approximation of what would be a dynamical Hooke's law. This viscoelastic approach is  the starting point of studies of mechanical dynamical deformations in materials \cite{truesdell_non-linear_2004}.

Finally, there is  the coupling between these two systems, labeled by ${\mathscr{L}}_\mathrm{sm}$ and commonly called ``magnetostriction'', in this context \cite{du_tremolet_de_lacheisserie_magnetostriction:_1993}.

More precisely, these Lagrangians are given by the expressions:
\begin{subeqnarray}
	\mathscr{L}_{\mathrm{s}}&=&\frac{m_s}{2}\dot{s}_i^2+\dot{s}_iA_i[\bm{s}]-V_s[\bm{s}]\slabel{Lagrangian-s}\\
	\mathscr{L}_{\mathrm{m}}&=&\frac{m_{\epsilon}}{2}\dot{\epsilon}_{ij}^2-V_\epsilon[\bm{\epsilon}]\slabel{Lagrangian-m}\\
	\mathscr{L}_{\mathrm{sm}}&=&-\frac{1}{2}B_{ijkl}\dot{s}_i\dot{s}_j\epsilon_{kl}
	\slabel{Lagrangian_coupling}
\end{subeqnarray}
These can be understood as describing   interacting objects. One is a point--like particle, whose position is labeled by $s_i(t)$. The other is, in fact, an extended object, whose ``position'' is  $\epsilon_{ij}(t)$. Latin indices run  from 1 to 3, and the Einstein summation convention of repeated indices is assumed.

The Lagrangian $\mathscr{L}_{\mathrm{s}}$ is invariant under local $U(1)$  transformations, i.e. $\delta A_i=\partial_i f(\bm{s})$, $\delta s_i=0$, since the Lagrangian changes by a total derivative \cite{soper_classical_2008}.

The first particle couples to the vector potential ${\bm A}[{\bm{s}}]$, which describes a physical magnetic field--however, since it is only magnetically charged, it  couples through its gyromagnetic ratio.

Because $\dot{\bm{s}}$ represents the spin variable, $m_s$ is an inertia constant which is here to describe  the precession and may be interpreted as a Land\'e factor, $V_s$ is a scalar potential, that gives rise to an  ``electric field" which can  affect  the conservation of the norm of the magnetization vector.
By pursuing  the analogy with  the charged particle in an electromagnetic field, $\bm{A}[\bm{s}]$ is a vector potential, which depends on the whole history of ${\bm{s}(t)}=\int_0^t\dot{\bm s}(\tau)d\tau$ and, as remarked above, transforms under $U(1)$.

The elastic medium is considered spatially uniform and the second Lagrangian describes the deformation of the elastic medium~\cite{dzyaloshinskii_poisson_1980}.
Eq.(\ref{Lagrangian-m}) means, in particular, that $\mathscr{L}_\mathrm{m}[\epsilon]$ defines a matrix model so the trace operation is implicitly assumed.
Moreover, if the elastic medium is isotropic, this term is invariant under local $SO(3)$ transformations, that act with the adjoint action: $\epsilon_{ij}\to \left[R\epsilon R^\mathrm{T}\right]_{ij}$, with $R\in SO(3)$;  so the full symmetry group of the theory, without interaction between particles, is $U(1)\times SO(3)$.

In the expression of ${\mathscr{L}}_{m}$, $m_\epsilon$ is an inertia term for the mechanical part. $V_\epsilon$ represents a scalar mechanical potential and can be expressed in an elastic medium as $V_\epsilon=\frac{1}{2}C_{ijkl}\epsilon_{ij}\epsilon_{kl}$ where $C$ is the elastic stiffness tensor.
Associated to this tensor, there is an elastic compliance tensor $S$ such that $C_{ijkl}S_{ijmn}=\frac{1}{2}\left(\delta_m^k\delta_n^l+\delta_n^k\delta_m^l\right)$.

Finally, for ${\mathscr{L}}_{\mathrm{sm}}$, $B_{ijkl}$ is a coupling matrix responsible for magnetostriction which is taken independent of the dynamical variables \cite{du_tremolet_de_lacheisserie_magnetostriction:_1993}.
For the interaction term to be, also, invariant under $U(1)\times SO(3)$, the fields, $s_i$ and $\epsilon_{ij}$ must carry ``charges'' that are related in a quite specific way~\cite{soper_classical_2008,consolo_lagrangian_2011}. In the case at hand, the invariance of the Lagrangian $\mathscr{L}_\mathrm{sm}=\dot{s}_i\dot{s}_j\mathrm{Tr}_{SO(3)}\left[B_{ijkl}\epsilon_{kl}\right]$ requires that $B$ transforms itself as $B_{ijkl}\to [R^TBR]_{ijkl}$ with the proper selections of indices.

In all these expressions the indices are ``space--like'' and an immediate question is, whether the rotational symmetry thus implied can be promoted to a full--fledged, emergent, Lorentz symmetry.
It is here that the ``no--interaction theorem''~\cite{leutwyler_no-interaction_1965} is relevant and implies that this is not possible, with a fixed--here two--number of particles (or for a matrix of fixed, finite, rank, referring to the $\epsilon_{ij}$).
This means, in particular, that, even if both inertia coefficients, $m_s$ and $m_\epsilon$, vanish, the excitations are not, in fact, massless, since the emergent Lorentz invariance is not compatible with any interaction term.  How Lorentz invariance can emerge in such systems is, currently, the subject of considerable activity--but the constraints from the no--interaction theorem seem not to have been fully appreciated and deserve further investigation. In the following we shall work out some of the consequences of the $U(1)\times SO(3)$ symmetry as acting on the spatial indices.

In order to probe the dynamics of all the internal system variables, external sources are necessary. These sources can--and here will be assumed to--couple minimally to the fields and give rise to force terms in the equations of motion.

For forces that can be expressed  in terms of scalar  potentials, we have $\mathscr{L}_{sources}=-j_i^{\mathrm{ext}}[{\bm s}]{\dot{s}}_i-\sigma_{ij}^{\mathrm{ext}}\epsilon_{ij}$.
At this step, regarding the magnetic part, ${\bm j}^{\mathrm{ext}}[\bm{s}]$ is a conserved current and cannot give rise to  a spin transfer torque (STT).
$\bm{\sigma}^{\mathrm{ext}}$ is an external, spatially uniform and instantaneous mechanical stress tensor.
Extensions to non-instantaneous and non-uniform sources do not present any conceptual difficulties \cite{goldstein_classical_2002}.

In order to derive expressions for the dissipative contribution in the Lagrangian formalism, one can remark  that Gilbert's dissipation functions for spins and STT can be mapped to currents, when they are not functions of ${\bm s}$ only, but also of higher order time derivatives such as :
\begin{equation}
\frac{\partial{\mathscr{L}_{losses}}}{\partial\ddot{s}_i}=\alpha\epsilon_{ijk}\dot{s}_j\ddot{s}_k+J(\dot{s}_i\dot{s}_jp_j-p_i\dot{s}_j\dot{s}_j)
\end{equation}
where $J$ is the amplitude of the current  and ${\bm p}$ its direction. As expected, the sign of the spin-torque dissipation function depends, apart from the direction of the current flow, on the relative magnetization configuration of the magnetic layers.

Using the same kind of reasoning, the elastic current $\sigma_{ij}$ can be decomposed into two terms
\begin{equation}
\sigma_{ij}=\sigma^{\textrm{ext}}_{ij}-\gamma\dot{\epsilon}_{ij}
\end{equation}
where $\sigma^{\textrm{ext}}_{ij}$ are the components of an external applied stress tensor, which derive from a potential energy function, and $\gamma$ is a mechanical damping constant, which is proportional to the strain time rate.

For each dynamical variable, Euler-Lagrange equations of motions (EOM)
\begin{subeqnarray}
\frac{\partial \mathscr{L}}{\partial s_i}-\frac{d}{dt}\left(\frac{\partial \mathscr{L}}{\partial \dot{s}_i}\right)&=&\frac{\partial\mathscr{L}_{sources}}{\partial \dot{s}_i}+\frac{\partial\mathscr{L}_{losses}}{\partial \ddot{s}_i}\\
\frac{\partial \mathscr{L}}{\partial \epsilon_{ij}}-\frac{d}{dt}\left(\frac{\partial \mathscr{L}}{\partial \dot{\epsilon}_{ij}}\right)&=&\frac{\partial\mathscr{L}_{sources}}{\partial\epsilon_{ij}}+\frac{\partial\mathscr{L}_{losses}}{\partial\dot{\epsilon}_{ij}}
\end{subeqnarray}
take the form
\begin{subeqnarray}
m_s\ddot{s}_i+F_{ij}\dot{s}_j+\frac{\partial V_s}{\partial s_i}-B_{ijkl}\left(\ddot{s}_j\epsilon_{kl}+\dot{s}_j\dot{\epsilon}_{kl}\right)&=j_i\slabel{eom-s}\\
m_\epsilon\ddot{\epsilon}_{ij}+\frac{\partial V_\epsilon}{\partial \epsilon_{ij}}+\frac{1}{2}B_{klij}\dot{s}_k\dot{s}_l&=\sigma_{ij}\slabel{eom-m}
\end{subeqnarray}
where the antisymmetric Faraday tensor $F$ is defined as usual~:
\begin{eqnarray*}
F_{ij}&=&\frac{\partial A_i}{\partial s_j}-\frac{\partial A_j}{\partial s_i}
\end{eqnarray*}
and describes  spin precession, since it can be mapped to a dual pseudovector $\bm{\omega}$
\begin{eqnarray*}
F_{ij}&\equiv&\epsilon_{ijk}\omega_{k}[{\bm s}].
\end{eqnarray*}
$\bm{\omega}$ is understood as the effective frequency of precession, and is usually defined as $\omega_i\equiv -\frac{1}{\hbar}\frac{\partial{H}}{\partial s_i}$, where $H$ is the total spin hamiltonian, whose precise expression depends on the nature of the considered magnetic interactions.

The current $j_i=j_i^{\mathrm{ext}}+\frac{\partial{\mathscr{L}_{losses}}}{\partial\ddot{s}_i}$ is then the total torque applied on the spin system.

In more conventional terms, the bulk  magnetization $\bm{M}(t)$, can be  identified with the vector $Ng\bm{\mu}(t)/V$, where $N$ is the number of magnetic moments, $V$ is the volume and $g\equiv m_s$ the Landé factor.
The magnetic induction $\bm{B}$ can be  identified with the expression
\begin{equation}
\label{Bfield}
\bm{B}=-\frac{1}{g\mu_B}\frac{\partial H}{\partial\bm{\mu}}
\end{equation}
with $\mu_B$ is the Bohr’s magneton.
Finally, the magnetic field $\bm{H}$ can be defined by the relation between the magnetic induction and the magnetization
\begin{equation}
\label{Hfield}
\bm{H}=-\bm{M}+\frac{\bm{B}}{\mu_0}
\end{equation}
with $\mu_0$ the permeability of the vacuum.

An advantage of our formulation is that these conventional quantities can be understood as emergent from a microscopic approach, that highlights the significance of the history of the sample. So in the following, we shall use the microscopic variables to describe the dynamics, since their relation to the conventional, macroscopic variables is transparent and allows a direct description of multisublattice effects, that have become of practical relevance and  are much harder to unravel in terms of the macroscopic variables.

For it has been demonstrated that, as in ferromagnets, in multisublattice magnetic systems, also,  the spin-polarized electrons transfer spin torques on each of the atomic sites \cite{nunez_theory_2006,xu_spin-transfer_2008,haney_ab_2007,haney_current-induced_2007}.
Consequently,  the magnetic structure of anti-ferromagnets (AFMs) may be described using  ``colored" vectors ${\bm{s}}^{L}$ and strain matrices $\epsilon_{ij}^{L}$, that arise due to strong exchange magnetic coupling, where $L$ labels the different inequivalent sites (or the sublattices).

 The  EOM take the form
\begin{subeqnarray}
\!\!\!\!m^{L}_{s}\ddot{s}^{L}_i+F_{ij}\dot{s}^{L}_j+\frac{\partial V_s}{\partial s^{L}_i}-B_{ijkl}\left(\ddot{s}^{L}_j\epsilon^{L}_{kl}+\dot{s}^{L}_j\dot{\epsilon}^{L}_{kl}\right)\!\!=j^{L}_i\slabel{eom-sJ}\\
\!\!\!\!m^{L}_{\epsilon}\ddot{\epsilon}^{L}_{ij}+\frac{\partial V_\epsilon}{\partial \epsilon^{L}_{ij}}+\frac{1}{2}B_{klij}\dot{s}^{L}_k\dot{s}^{L}_l\!\!=\sigma^{L}_{ij}\slabel{eom-mJ}
\end{subeqnarray}
where $j_i^L=j_i^{\mathrm{ext}}+\alpha\epsilon_{ijk}\dot{s}^L_j\ddot{s}^L_k+J(\dot{s}^L_i\dot{s}^L_jp_j-p_i\dot{s}^L_j\dot{s}^L_j)$ and $\sigma_{ij}^L=\sigma_{ij}^{\textrm{ext}}-\gamma\dot{\epsilon}_{ij}^L$.

Since the  variable we are, really,  interested in is ${\bm \mu}(t)\equiv\dot{{\bm s}}(t)$,  we can rewrite the system as
\begin{subeqnarray}
	\!\!\!\!m^{L}_{s}\dot{\mu}^{L}_i+F_{ij}\mu^{L}_j+\frac{\partial V_s}{\partial s^{L}_i}-B_{ijkl}\left(\dot{\mu}^{L}_j\epsilon^{L}_{kl}+\mu^{L}_j\dot{\epsilon}^{L}_{kl}\right)\!\!=j^{L}_i,\slabel{eom-sJL}\\
	\!\!\!\!m^{L}_{\epsilon}\ddot{\epsilon}^{L}_{ij}+\frac{\partial V_\epsilon}{\partial \epsilon^{L}_{ij}}+\frac{1}{2}B_{klij}\mu^{L}_k\mu^{L}_l\!\!=\sigma^{L}_{ij}.\slabel{eom-mJL}
\end{subeqnarray}
In the absence of any mechanical damping and inertia, only in the single lattice site situation $L=1$, does eq.(\ref{eom-mJL}) lead to  the well-known result that $\epsilon_{ij}$ is  the sum of the Hooke's law contribution and that of a tensor, which is   a quadratic function of the magnetization \cite{du_tremolet_de_lacheisserie_magnetostriction:_1993}, i.e.~:
\begin{equation}
	\epsilon_{ij}=S_{ijkl}\sigma^{\rm ext}_{kl}-\frac{1}{2}S_{ijkl}B_{klmn}\mu_m\mu_n,
\end{equation}
Whereupon any reference to the site $L$ can be safely dropped.

\section{L=2 antiferromagnetic switching}
We now proceed to the simplest sublattice case, where $L=2$--an AFM  with two magnetic sublattices. We want to focus on the phenomenon of ``switching'', which is the reversal of the magnetization on {\em both} sublattices.

We start by  studying  the case  where there is no spin potential acting on the  lattice sites, that can affect  the norm of the magnetization  i.e. ${\displaystyle\frac{\partial V_s}{\partial s^{L}_i}=0}$. Moreover there is no external spin current, i.e. ${\bm j}^{\mathrm{ext}}={\bm 0}$.

In the particular case of an AFM with two magnetic sublattices, it is useful to define the net magnetization ${\bm m}\equiv\frac{1}{2}\left({\bm \mu}^{1}+{\bm \mu}^{2}\right)$ and the Néel order parameter ${\bm l}\equiv \frac{1}{2}\left({\bm \mu}^1-{\bm \mu}^2\right)$~\cite{roepke_derivation_1971,gomonay_spin_2010,cheng_spin_2014,cheng_ultrafast_2015}, as well as  the  corresponding strain matrices (FM strain) $\epsilon_{ij}\equiv \frac{1}{2}\left(\epsilon_{ij}^{1}+\epsilon_{ij}^{2}\right)$ and (AFM strain) $\eta_{ij}\equiv\frac{1}{2}\left(\epsilon_{ij}^{1}-\epsilon_{ij}^{2}\right)$.

Eqs.(\ref{eom-sJ}) can then be  reformulated in terms of  a mass matrix $M^L$ and an effective Faraday matrix $D^L$, (which is not fully antisymmetric),  consistent with  Landau-Lifshitz-Gilbert-Slonczewski, i.e.~:
\begin{equation}
	M_{ij}^L\dot{\mu}_j^L+D_{ij}^L\mu^{L}_j=j^{L}_i
\end{equation}
with $M_{ij}^L\equiv\delta_{ij}m_s^L-B_{ijkl}\epsilon_{kl}^L$ and $D_{ij}^L\equiv F_{ij}-B_{ijkl}\dot{\epsilon}^{L}_{kl}$.
If $B$ is totally  antisymmetric, then $D_{ij}^L$ is, too, and when $\bm{j}={\bm 0}$, we recover the usual spin precession equation.

If the medium is isotropic, then both $C$ and $B$ can be expressed in terms of only two  independent material constants~:
\begin{subeqnarray*}
	B_{ijkl}&=&B_0\delta_{ij}\delta_{kl}+B_1(\delta_{ik}\delta_{jl}+\delta_{il}\delta_{jk})\\
	C_{ijkl}&=&C_0\delta_{ij}\delta_{kl}+C_1(\delta_{ik}\delta_{jl}+\delta_{il}\delta_{jk})
\end{subeqnarray*}
For practical reasons, these material constants $B_0$, $B_1$, $C_0$ and $C_1$ are dimensionless by dividing out $\mu_0M_s^2$ where $M_s$ is the saturation magnetization. Incidentally, we also divide the applied external stress $\sigma_{ij}^\mathrm{ext}$ by the same factor, to produce a dimensionless stress.

Eqs.(\ref{eom-s},\ref{eom-m}) are integrated using a Runge-Kutta (RK) numerical scheme of order 4-5 with a variable integration timestep.

In order to check the validity of this integration scheme, we used a RK scheme of higher order and we did not observe any differences between the results.
To address longer simulation times or systems with larger sites, an extensive study would imply using a better numerical integrator, with symplectic structure, for conservation,  not only of  the phase space volume, but also of the structure of the system of equations. This will be discussed  in future work.

\begin{figure}[htp]
	\centering
	\resizebox{0.98\columnwidth}{!}{\includegraphics{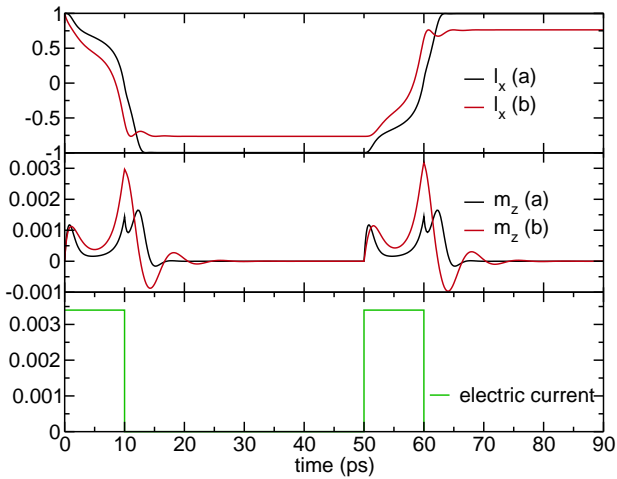}}
	\caption{
		Out-of-plane magnetization $m_z$ and staggered field projected on the easy asis as functions of time. \{$m_s=1$, $m_\epsilon=0$, $C_0=5.1\times 10^5$, $C_1=3.5\times 10^5$, $B_0=7.7$, $B_1=-23$, $\sigma^{\mathrm{ext(a)}}_{11}=\sigma^{\mathrm{ext(a)}}_{22}=\sigma^{\mathrm{ext(a)}}_{33}=100$, $\gamma=2\times 10^6$, $\alpha=0.005$, $J=0.0034$rad.THz, $\omega_a=2\pi$rad.GHz, $\omega_E$=172.16 rad.THz, $M_s=5.10^5\mathrm{A.m}^{-1}$\}. Initial conditions: ${\bm s}_1(0)=-{\bm s}_2(0)=\hat{\bm x}$. (a) is indistinguishable from the uncoupled situation and (b) is $\sigma^{\mathrm{ext(b)}}_{11}=\sigma^{\mathrm{ext(b)}}_{22}=\sigma^{\mathrm{ext(b)}}_{33}=3\times10^4.$ \label{Fig1}}
\end{figure}

It has been recently reported~\cite{cheng_ultrafast_2015} that polycrystalline NiO is a candidate for antiferromagnetic switching.
Upon neglecting, at first, magnetostrictive terms,  (i.e. setting $B_0=B_1=0$), we find that our model where such a  material is described by $L=2$ spins, interacting only through an anti-ferromagnetic exchange coupling, corresponding to a precession frequency, $\omega_E$, is in perfect agreement with these results.
This situation is reported in Fig.~\ref{Fig1}, labeled by (a).

To induce such switchings,  the external current is taken as a stream of square electrical pulses, along the $\hat{\bm z}$-axis ($\bm{p}=\hat{\bm{z}}$). In addition to the exchange interaction, $\omega_E$, the spins are subject to a (global) anisotropy, of strength $\omega_a$, along the $\bm{x}-$axis and the ``spin accumulation'', $m_z$ is monitored.

Once these results are available, it is possible to study more general situations, namely to check for time symmetric behavior of such a system, by introducing a second electrical pulse, that should bring the spin system back to its original state.

In order to prove that the system is, indeed, symmetric under time reversal, taking into account, in particular, the mechanical stresses, we consider that an isotropic pressure $P$ is applied, i.e. $\sigma_{ii}=P/3$ $\forall i\in 1,2,3$. For Fig.~\ref{Fig1}, we have $P^{\mathrm{(a)}}\approx0.1\mathrm{GPa}$ and $P^{\mathrm{(b)}}\approx30\mathrm{GPa}$. The largest value of P was selected to display more clearly the effect of the coupling on the magnetic system.

Moreover, the conditions and values of the simulation are identical to those found in reference \cite{cheng_ultrafast_2015}.

We start the simulations using an initial configuration, where spins are aligned along the $\bm{x}$-axis in an antiferromagnetic situation and apply two electric pulses separated by 50 ps.

The profiles of the Néel vector and of the spin accumulation, when taking into account magnetostrictive effects, are displayed in Fig.~\ref{Fig1}, curves (b)--to be compared to curves (a), where the spin--lattice coupling is zero--and to the results of reference~\cite{cheng_ultrafast_2015}.

As already observed, because the STT acts as a strong damping, the $\hat{\bm z}$-component of the spin vector is slightly shifted from a purely anti-parallel situation during the pulses and the whole system reverses spin orientation, as shown in Fig.~\ref{Fig1} by checking the value taken by the Néel vector ${\bm l}$. However for the expanded sample $(b)$, the switching rate seems to be faster and the spin accumulation appears larger. Intuitively one would rather think that a compression would enhance exchange interactions in the material and hence lead to faster switching rates, which seems not to be the case.

We use references \cite{grimvall_thermophysical_1999,plessis_elastic_1971,plessis_magnetostriction_1971} to get numerical values for $C_0$, $C_1$ and the traditional magnetostriction coefficients $\lambda_s$ along known directions are obtained. These coefficients are defined as $\lambda_s=\beta_i\epsilon_{ij}\beta_j$ where $\bm{\beta}$ is the unit vector along which the deformation is projected.

In order to obtain the magneto-elastic constants $B_0$ and $B_1$, an inversion formula is needed. If the effect of the magneto-elastic constants only is considered, then at equilibrium, the strain tensor can be calculated. As a result one obtains
\begin{equation}
\epsilon^{\mathrm{eq}}_{ij}\approx\frac{\frac{1}{2}\left(\frac{C_0B_1}{C_1}-B_0\right)}{3C_0+2C_1}\delta_{ij}-\frac{B_1}{2C_1}\mu^{\mathrm{eq}}_i\mu^\mathrm{eq}_j
\label{eqstrain}
\end{equation}
which corresponds to the tensorial expression of the equilibrium magnetostriction when the sample is magnetically saturated  along a chosen direction, here for example $\bm{x}$ (i.e. ${\mu}^{\mathrm{eq}}_x=1$ and ${\mu}^{\mathrm{eq}}_y={\mu}^{\mathrm{eq}}_z=0$ ). As our material displays spherical symmetry, we can choose any axis, thus we chose one of the simplest situation. Now the measured quantity actually is the projection of this equilibrium deformation along a given vector $\bm{\beta}$. Thanks to references~\cite{grimvall_thermophysical_1999,plessis_elastic_1971,plessis_magnetostriction_1971} we have experimental data for the magnetostriction along the $\bm{x}$-axis which we shall call longitudinal, denoted by $\lambda_s^L$ and the striction along the $\bm{y}$-axis (or any axis in the ($\bm{y}$, $\bm{z}$) plane for this matter) which we shall call transverse, denoted by $\lambda_s^T$. This gives us the following expressions to find $B_0$ and $B_1$
\begin{eqnarray}
	\lambda_s^L&=&\epsilon^{\mathrm{eq}}_{xx}\approx\frac{\frac{1}{2}\left(\frac{C_0B_1}{C_1}-B_0\right)}{3C_0+2C_1}-\frac{B_1}{2C_1}\\
	\lambda_s^T&=&\lambda_s^L+\frac{B_1}{2C_1}
\end{eqnarray}

Without any magneto-elastic coupling, when the mechanical system is subject to constant external stress only, the values of its strain at equilibrium are given by
\begin{equation}
\label{straineq0}
\epsilon^\mathrm{eq}_{ij}=\sigma^{\mathrm{ext}}_{ij}/(3C_0+2C_1)
\end{equation}
Because the values of the magneto-elastic constants $B$ are, typically, $10^{5}$ times smaller than the mechanical constants $C$, the dynamical effect of the spin on the mechanical system can be estimated  by first solving the spin dynamics without any coupling and considering the contribution of the $B$s, displayed in eq.~(\ref{eqstrain}) as a perturbation.

It must be kept in mind that this approximation works only as long as the system stays close to the mechanical equilibrium.
One can indeed check that the numerical values at equilibrium are consistent  with the results given by eq.(\ref{eqstrain}).

\begin{figure}[htp]
	\centering
	\resizebox{0.98\columnwidth}{!}{\includegraphics{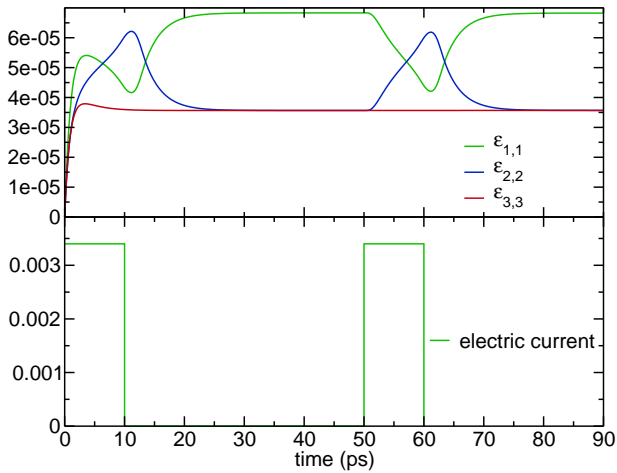}}
	\caption{
	Diagonal strain components as functions of time. Conditions are identical than Fig.~\ref{Fig1}(a).\label{Fig2}}
\end{figure}

NiO polycrystals are not known to be highly magnetostrictive materials and the computed magneto-elastic constants are small.
Thus we observe that the mechanical response is much more sensitive to the coupling with the magnetic degrees of freedom than the other way around as it is shown in Fig.~\ref{Fig2}.
Under the same STT stream of pulses, we plot the diagonal components of the strain in Fig.~\ref{Fig2}, using significantly lower values for the external stress, than those depicted in Fig.~\ref{Fig1}.(b).

More precisely, the response one gets from the mechanical system in reaction to a magnetical stimulus is more significant than the reaction from the magnetic part to an external stress.

One can indeed notice that the mechanical equilibrium deformation is displaced according to the relations, deduced from eqs.(\ref{eqstrain}) but also that the dynamics  clearly shows the effects  of the SST pulses.

On the other hand, with increasing external stress, one notices that the sensitivity of the mechanical response to the coupling with the magnetic degrees of freedom is blunted.

\begin{figure}[htp]
	\centering
	\resizebox{0.98\columnwidth}{!}{\includegraphics{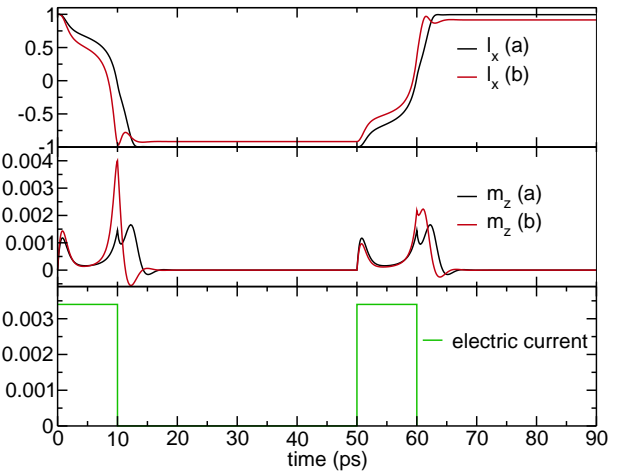}}
	\caption{
		Out-of-plane magnetization $m_z$ and staggered field projected on the easy asis as functions of time. Numerical constants are identical to Fig.~\ref{Fig1} except for the external stress where non zero components are $\sigma^{\mathrm{ext}(a)}_{12}=\sigma^{\mathrm{ext}(a)}_{21}=100$ for (a) and $\sigma^{\mathrm{ext}(b)}_{12}=\sigma^{\mathrm{ext}(b)}_{21}=5000$ for (b).
\label{Fig3}}
\end{figure}

In the case of an external shear instead of a tensile stress, the effect  on the switching seems to appear, already,  at weaker external stresses, as  shown in Fig.~\ref{Fig3}.

It seems to be possible to recover a switching state for a stress six times weaker than that for  pure tensile stress.
The side effect is asymmetry in the  magnetic switching dynamics.
Indeed, the spin accumulation along the $\hat{\bm z}$ direction is modified as the mechanical state  changes between the two pulses.
For the first pulse, the mechanical system is still relaxing towards equilibrium,  whereas around the second pulse, it has already attained a new equilibrium state, that produces an asymmetric switch.

In order to exhibit measurable backreactions from both mechanical to magnetical systems in NiO polycrystals, we need to consider very large applied stresses, not easily obtained  in real experiments.

However, one can think of  insulator materials in multilayers that may present larger  magneto-mechanical coupling constants\cite{koon_giant_1991,clark_magnetostrictive_2002}, which would significantly lower the external stress values that would  be needed to produce comparable effects.

\section{Discussion}
In this paper, we have investigated numerically several aspects of the dynamics of the spin-lattice coupling that describes magnetostrictive effects. We have used a microscopic approach for defining magnetic and elastic degrees of freedom, in terms of which the conventionally used, macroscopic quantities can be understood as emergent. Our approach leads to the identification of novel symmetries, whose experimental consequences can be studied in detail with current and future technology and can lead to new insights for theoretical and computational models.

We have considered a fixed lattice of size $L$, for which each site carries the physical degrees of freedom, that pertain to the actual time-evolution of the system, namely magnetic moments and elastic deformations.
Thus the underlying change in the magnetic response,  due to the external stress, is taken into account through an effective coupling term, whose form is largely determined by the symmetries of the problem.

However the numerical values of the parameters must be determined by a molecular dynamics model, that relies on ``moving particle strategies'',   that describe in microscopic detail,   the intensity and direction of the magnetic atomic interaction, as functions of the distance between atoms \cite{beaujouan_anisotropic_2012,beaujouan_thermal_2012}.
The mesoscopic approach developed in the present study thus complements the magnetic molecular dynamics and provides  a multiscale framework for  computing  both magnetic and mechanical properties of materials.
Moreover, additional baths, whether thermal, quantum or due to disorder,   can be readily taken int account  at this level of modeling, since  eqs.(\ref{eom-sJL}) and (\ref{eom-mJL}) may be modified consistently, along the lines of  \cite{thibaudeau_nambu_2017}. The details will be presented in forthcoming work.
\bibliographystyle{unsrt}

\end{document}